\documentstyle[12pt]{article}
\def\Re{{\cal R \mskip-4mu \lower.1ex \hbox{\it e}\,}}
\def\Im{{\cal I \mskip-5mu \lower.1ex \hbox{\it m}\,}}
\def\ie{{\it i.e.}}
\def\eg{{\it e.g.}}

\def\etal{{\it et al.}}

\def\sub#1{_{\lower.25ex\hbox{$\scriptstyle#1$}}}
\def\sul#1{_{\kern-.1em#1}}
\def\sll#1{_{\kern-.2em#1}}
\def\sbl#1{_{\kern-.1em\lower.25ex\hbox{$\scriptstyle#1$}}}
\def\ssb#1{_{\lower.25ex\hbox{$\scriptscriptstyle#1$}}}
\def\sbb#1{_{\lower.4ex\hbox{$\scriptstyle#1$}}}

\def\gev{\,{\rm GeV}}

\def\to{\rightarrow}
\def\mh{\ifmmode m\sbl H \else $m\sbl H$\fi}
\def\mch{\ifmmode m_{H^\pm} \else $m_{H^\pm}$\fi}
\def\mt{\ifmmode m_t\else $m_t$\fi}
\def\mc{\ifmmode m_c\else $m_c$\fi}
\def\mz{\ifmmode M_Z\else $M_Z$\fi}
\def\mw{\ifmmode M_W\else $M_W$\fi}
\def\mws{\ifmmode M_W^2 \else $M_W^2$\fi}
\def\mhs{\ifmmode m_H^2 \else $m_H^2$\fi}
\def\mzs{\ifmmode M_Z^2 \else $M_Z^2$\fi}
\def\mts{\ifmmode m_t^2 \else $m_t^2$\fi}
\def\mcs{\ifmmode m_c^2 \else $m_c^2$\fi}
\def\mchs{\ifmmode m_{H^\pm}^2 \else $m_{H^\pm}^2$\fi}
\def\ztwo{\ifmmode Z_2\else $Z_2$\fi}
\def\zone{\ifmmode Z_1\else $Z_1$\fi}
\def\mtwo{\ifmmode M_2\else $M_2$\fi}
\def\mone{\ifmmode M_1\else $M_1$\fi}
\def\tb{\ifmmode \tan\beta \else $\tan\beta$\fi}
\def\xw{\ifmmode x\sub w\else $x\sub w$\fi}
\def\ch{\ifmmode H^\pm \else $H^\pm$\fi}
\def\lum{\ifmmode {\cal L}\else ${\cal L}$\fi}
\def\inpb{\ifmmode {\rm pb}^{-1}\else ${\rm pb}^{-1}$\fi}
\def\infb{\ifmmode {\rm fb}^{-1}\else ${\rm fb}^{-1}$\fi}
\def\epem{\ifmmode e^+e^-\else $e^+e^-$\fi}
\def\ppb{\ifmmode \bar pp\else $\bar pp$\fi}

\newskip\zatskip \zatskip=0pt plus0pt minus0pt
\def\matth{\mathsurround=0pt}

\def\atversim#1#2{\lower0.7ex\vbox{\baselineskip\zatskip\lineskip\zatskip
  \lineskiplimit 0pt\ialign{$\matth#1\hfil##\hfil$\crcr#2\crcr\sim\crcr}}}

\renewcommand{\thefootnote}{\fnsymbol{footnote}}

\begin{document} \begin{titlepage}
\setcounter{page}{1}
\thispagestyle{empty}
\rightline{\vbox{\halign{&#\hfil\cr
&ANL-HEP-PR-93-19\cr
&April 1993\cr}}}
\vspace{1in}
\begin{center}

{\Large\bf
Constraints on Anomalous Gauge Boson Couplings From $b \to s\gamma$}
\footnote{Research supported by the
U.S. Department of
Energy, Division of High Energy Physics, Contracts W-31-109-ENG-38.}
\medskip

\normalsize THOMAS G. RIZZO
\\ \smallskip
High Energy Physics Division\\Argonne National
Laboratory\\Argonne, IL 60439\\

\end{center}

\begin{abstract}

The recent results of the CLEO Collaboration on both inclusive and exclusive
radiative $B$ decays are combined with those of the UA2 Collaboration
on $W\gamma$ production to highly constrain the anomalous trilinear gauge
couplings of the $W$. The theoretical analysis of the $b \to s\gamma$ process
employs, next-to-leading order operator coefficient evolution as well as QCD
bremsstrahlung and appropriate phase space corrections.

\end{abstract}



\renewcommand{\thefootnote}{\arabic{footnote}} \end{titlepage}


The Standard Model(SM) provides an excellent description of presently
available experimental data{\cite {review}}. However, it is believed that the
SM cannot be the whole story as it leaves too many questions unanswered.
Whatever new physics actually exists beyond the SM may be more subtle in its
first appearance than the production of an unexpected particle at a hadron or
$e^+e^-$ collider. Thus, one approach in looking for new physics is to probe
for small deviations from the predictions of the SM either through the detailed
analysis of high precision data or by looking for rare processes which are
either highly suppressed or forbidden in the SM.

One aspect of the SM which has only recently begun to be directly
tested{\cite {ua2}},
is the trilinear gauge coupling of the photon or $Z$ to $W^+W^-$. Models
which introduce `anomalous' couplings are convenient test beds for looking
for deviations from the SM and tell us something about `how the SM is doing.'
There have been many discussions about such couplings in the literature{\cite
{bigref}} many of which have recently been called into question {\cite {der}}.
Such analyses take either one of two approaches: looking for deviations from
the SM via tree-level processes, such as $e^+e^- \to W^+W^-$ at LEP II or
$p \bar p \to W\gamma$ at the Tevatron, or they examine the influence of the
anomalous couplings on loop-order processes. The latter procedure can be
particularly dangerous as in many cases, such as the $g-2$ of the muon,
cutoffs must be introduced to regulate loop integrals and can err in
attributing a physical significance(\eg , the scale of new physics) to the
cutoff. This point has recently been stressed by Burgess and London
{\cite {bur}}.

Some loop-order processes do {\it {not}} suffer from this difficulty, due to
the cancellations provided by
the GIM mechanism{\cite {GIM}}, and can yield cutoff-independent bounds on
any anomalous couplings. One example of such a process is the reaction $b \to
s\gamma$ which we will examine in the discussion below. Assuming that CP is
conserved, the usual analysis of the $WW\gamma$ trilinear coupling postulates
the existence of only two additional parameters: $\kappa =1+\Delta\kappa$ and
$\lambda$, with the SM limit being $\Delta\kappa, \lambda=0$. Analyses of the
$b \to s\gamma$ process including the effects of either of these new couplings
{\it {separately}} to leading order in the QCD corrections already exist in
the literature{\cite {chia,peterson}}. Our philosophy will be that since we
are quite ignorant of what new physics may lie beyond the SM, the size of
any anomalous couplings should be treated as {\it {a priori}} unknowns and
that we will let experimental data tell us what the bounds on these
parameters are.

In the present paper we perform an
analysis of this reaction to study the effects of both anomalous couplings
{\it {simultaneously}} including next-to-leading order QCD as well as other
corrections. In addition, a {\cite {new}} new upper bound on the inclusive
$b \to s\gamma$ process now exists from
CLEO: $B(b \to s\gamma)<5.4\times 10^{-4}$, which further strengthens the
constraints we obtain. (The actual observation of the $B \to K^* \gamma$ mode
also provides a new {\it {lower}} limit to the inclusive rate provided some
rather weak theoretical assumptions are made.) Finally, our results from the
$b \to s\gamma$ process are combined with the limits from the
UA2 Collaboration showing that a only a `relatively small' region of the
$\Delta\kappa-\lambda$ plane remains allowed.

Our analysis proceeds as follows. In order to calculate the inclusive $b \to
s\gamma$ branching fraction we begin as usual by scaling our expression for
the $b \to s\gamma$ rate to the corresponding
theoretical prediction for the semileptonic decay rate. (This removes a major
uncertainty in the calculation associated with the overall factor of the
fifth power of the
b-quark mass appearing in both expressions.) We then use the latest data
on the semileptonic branching fraction{\cite {pdg,drell}} to rescale our
result, \ie, $B(b \to X \ell \nu)=0.108$:
\begin {equation}
B(b \to s\gamma) = {\Gamma(b \to s\gamma)\over {\Gamma(b \to X\ell \nu)}}
B(b \to X\ell \nu)
\end {equation}
The semileptonic rate is calculated including both charm and non-charm modes,
assuming $V_{ub}/V_{cb}=0.1$, and includes both phase space and QCD corrections
with $m_b=5\gev$ and $m_c=1.5\gev${\cite {cab}}. We note the important
observation that a choice of a smaller value of $m_b$ will result in even
stronger bounds than the ones presented below since the SM prediction for the
$b \to s\gamma$ decay rate increases as the b-quark mass decreases. The
calculation of the
$b \to s\gamma$ rate itself uses the next-to-leading log evolution equations
for the coefficients of the operators in the effective Hamiltonian due to
Misiak{\cite {misiak}}, the gluon bremsstrahlung corrections of Ali and
Greub{\cite {ali}}, the $m_{top} \neq M_W$ corrections of Cho and Grinstein
{\cite {cho}}, a running $\alpha_{QED}$ evaluated at the b-quark mass scale,
and 3-loop evolution of the running $\alpha_s$ matched to the value obtained
at the $Z$ scale via a global analysis{\cite {ellis}} of all data. Phase space
corrections for the strange quark mass in the final state were included and the
ratio of Kobayashi-Maskawa mixing matrix elements, $V_{tb}V_{ts}/V_{cb}$, was
assumed to have unit magnitude. The details of this procedure for the SM will
be presented elsewhere{\cite {jlh}}. To complete
the calculation we need to use the one-loop matching conditions for the
various operators{\cite {misiak}} in a form that includes contributions from
both the SM as well as the anomalous trilinear couplings. In practise, only the
 coefficient of the electromagnetic dipole transition operator, traditionally
called $O_7$, is modified by the presence of such additional terms. (The
coefficients of two other operators which do not mix with $O_7$ are also
modified.)
Symbolically, we can write the coefficient of this operator at the $W$ scale
in  the form $c_7(M_W)=c_7(M_W)^{SM}+\Delta\kappa A_1+\lambda A_2$, where{\cite
{chia,peterson}}
\begin{eqnarray}
c_7(M_W)^{SM} &=& -{1\over 2} \left[ {-3x^3+2x^2\over 2(1-x)^4}\ln x
-{8x^3+5x^2-7x\over 12(1-x)^3} \right] \nonumber \\
A_1 &=& -{1\over 4} \left[ {2x\over (1-x)^2} + {x^2(3-x)\over (1-x)^3}\ln x
\right] \\
A_2 &=& -{1\over 4} \left[ {x(1+x)\over (1-x)^2} + {x^2\over (1-x)^3}\ln x
\right] \nonumber
\end{eqnarray}
with $x=m_t^2/M_W^2$ as the only free parameter. Fig.~1 shows the {\it
{separate}} $\Delta\kappa$ and $\lambda$ dependencies of the $b \to s\gamma$
branching, $B$, for a few choices of $m_t$ which follow from this procedure.
Note that for most values of these
parameters, $B$ is quite large, \eg , $\geq 10^{-3}$, and would thus already
be in conflict with the old CLEO limit of $8.4\times 10^{-4}${\cite {old}}.
Fig.~2
shows the region in the $\Delta\kappa-\lambda$ plane which would be allowed as
this bound is tightened up and combined with those from the UA2{\cite {ua2}}
analysis. We see that as $m_t$ increases, the allowed region of parameter
space, given a specific bound on $B$, slowly shrinks.  The reason for this is
clear; as $m_t$ increases the SM prediction slowly begins to saturate the new
CLEO limit thus forcing a further restriction on non-SM physics.

Taking the new CLEO limit at face value, we show in Fig.~3 the $m_t$
dependence of the resulting allowed region of the $\Delta\kappa -\lambda$
parameter space. We must keep in mind when examining this figure the recently
improved lower bounds on $m_t$ announced by the CDF ($m_t > 108\gev$) and D0
($m_t>103\gev$) Collaborations{\cite {cdf,d0}}. These lower limits are
expected to increase further to the neighborhood of 120 GeV in the next few
months(provided the top is not found!).

How might the bounds on $\Delta\kappa$ and $\lambda$ be improved in the near
future, \ie , before the advent of LEP II? Several possibilities are likely:
($i$) Both CDF and D0 are expected to probe top masses up to the 150 GeV
region during the next Tevatron collider run. If top is found it will remove a
large uncertainty in our calculation; if not, the bounds on both the anomalous
coupling parameters will improve anyway as shown in Fig.~3. ($ii$) The actual
new CLEO upper limit on the branching fraction for $b \to s\gamma$
may eventually become
stronger than the result we use here as more data is accumulated. ($iii$) The
CDF Collaboration (and
perhaps later the D0 Collaboration) is expected to announce the results of
their analysis of the $p \bar p \to W\gamma$ process{\cite {larryuli}},
paralleling that of UA2, in the near future. One might expect that using the
data from the 1988-9 Tevatron run alone, CDF may be able to reduce the
size of the allowed
region in the $\Delta\kappa-\lambda$ plane by about a factor of 2{\cite {ed}}.
Of course, with the increased integrated luminosity accumulated from the
1992-3 run, this reduction in the allowed region might be somewhat larger.
($iv$)CLEO can conclusively
observe the $b \to s\gamma$ process. This is most likely to occur
through the observation of a particular exclusive mode, such as
$B \to K^*\gamma$, for which a reported upper limit of $9.2\times
10^{-5}${\cite {old}} has been known for some time. In fact, an actual
branching fraction(not a limit!) for the $B \to K^*\gamma$ process has just
recently been reported by CLEO: $B(B \to K^*\gamma)=(4.5\pm 1.5\pm 0.9)\times
10^{-5}${\cite {new}}.
We can, of course, safely conclude that the branching fraction for the
{\it {inclusive}} $b \to s\gamma$ process is larger than this value
since the ratio of the `exclusive-to-inclusive' branching fractions must be
less than unity. In fact, we can do substantially better
by noting that this `exclusive-to-inclusive' ratio is conservatively
expected to be less
than about 0.33{\cite {isgur,soni}}. If we take for purposes of demonstration
the value of $5\times 10^{-5}$ as the new lower limit on the inclusive rate
which, results from using
the factor of 0.33 as the `exclusive-to-inclusive' ratio, we see from Fig.~4
that a sizeable portion of the $\Delta \kappa -\lambda$ parameter space that
was allowed previously
would then be eliminated. Clearly, taken together, ($i-iv$) above are allowing
us to `home-in' on a rather small region of the parameter space which contains
the SM.

In this paper we have shown that by combining the new CLEO results on
radiative $B$ decays, a new theoretical calculation of the expected
branching fraction for such processes, and the analysis by UA2 of $W\gamma$
production puts strong constraints on the anomalous gauge boson couplings of
the $W$. We anticipate that further restrictions in the anomalous coupling
parameters will follow from ($i$)-($iv$) above during the next year. While
many believe both $\Delta \kappa$ and $\lambda$ must be small, it is wise
to have this confirmed by experiment.

\vskip.25in
\centerline{ACKNOWLEDGEMENTS}

The author would like to thank J.L.\ Hewett, E.\ Thorndike, and N.\ Deshpande
for discussions related to this work. The author would also like to thank
E.\ Thorndike for discussions about the new CLEO results. This research
was supported in part by the U.S.~Department of Energy
under contract W-31-109-ENG-38.

\newpage

%
\def\MPL #1 #2 #3 {Mod.~Phys.~Lett.~{\bf#1},\ #2 (#3)}
\def\NPB #1 #2 #3 {Nucl.~Phys.~{\bf#1},\ #2 (#3)}
\def\PLB #1 #2 #3 {Phys.~Lett.~{\bf#1},\ #2 (#3)}
\def\PR #1 #2 #3 {Phys.~Rep.~{\bf#1},\ #2 (#3)}
\def\PRD #1 #2 #3 {Phys.~Rev.~{\bf#1},\ #2 (#3)}
\def\PRL #1 #2 #3 {Phys.~Rev.~Lett.~{\bf#1},\ #2 (#3)}
\def\RMP #1 #2 #3 {Rev.~Mod.~Phys.~{\bf#1},\ #2 (#3)}
\def\ZP #1 #2 #3 {Z.~Phys.~{\bf#1},\ #2 (#3)}
\def\IJMP #1 #2 #3 {Int.~J.~Mod.~Phys.~{\bf#1},\ #2 (#3)}

\newpage

%
{\bf Figure Captions}
\begin{itemize}

\item[Figure 1.]{The branching fraction for $b \to s\gamma$ as a function of
(a)$\Delta \kappa$ and (b)$\lambda$, with the second anomalous coupling set to
zero. The dotted(dashed, dash-dotted, solid) curve corresponds to a top-quark
mass of 90(120, 150, 180) GeV.}
\item[Figure 2.]{Allowed region in the $\Delta \kappa-\lambda$ plane from the
analysis of UA2(inside the almost horizontal lines) and from $b \to s\gamma$
(between pairs of lines of the same type) assuming an upper limit on the
branching fraction of 8(7, 6, 5, 4, 3)$\times 10^{-4}$ corresponding to the
dotted(
dashed, dash-dotted, solid, squared-dotted)curve for a top-quark mass of
(a)120, (b)150, or (c)180 GeV.}
\item[Figure 3.]{Excluded region as in Fig. 2 assuming an upper limit of
$5.4\times 10^{-4}$ on the $b \to s\gamma$ branching fraction for $m_t=$ 90
(dotted), 120(dashed), 150(dash-dotted), 180(solid), or 210(square-dotted)
GeV.}
\item[Figure 4.]{Excluded region as in Fig. 3 for a top-quark mass of 150
GeV assuming an upper limit of $5.4\times 10^{-4}$ on the branching fraction
for the inclusive $b \to s\gamma$
process {\it {and}} a lower limit of $0.5\times 10^{-4}$ from the exclusive
$B \to K^*\gamma$ process using a ratio of the `exclusive-to-inclusive' ratio
of 0.33. The remaining allowed region lies between each pair
of dotted and solid lines subject to the bounds from UA2. `S' locates the
prediction of the SM.}
\end{itemize}

\end{document}